\documentclass[prc,showpacs,preprintnumbers,superscriptaddress,floatfix]{revtex4}
\usepackage{amsfonts}
\usepackage{graphicx}

\begin{document}

\title{Transmission resonances for a Dirac particle in a one-dimensional Hulth\'en potential}
\author{Jian You Guo}
\email{jianyou@ahu.edu.cn}
\affiliation{School of physics and material science, Anhui university, Hefei 230039, P.R.
China}
\author{Shao Wei Jin}
\affiliation{School of physics and material science, Anhui university, Hefei 230039, P.R.
China}
\author{Fu Xin Xu}
\affiliation{School of physics and material science, Anhui
university, Hefei 230039, P.R. China}
\begin{abstract}
We have solved exactly the two-component Dirac equation in the
presence of a spatially one-dimensional Hulth\'en potential, and
presented the Dirac spinors of scattering states in terms of
hypergeometric functions. We have calculated the reflection and
transmission coefficients by the matching conditions on the
wavefunctions, and investigated the condition for the existence of
transmission resonances. Furthermore, we have demonstrated how the
transmission-resonance condition depends on the shape of the
potential.
\end{abstract}

\pacs{03.65.Nk, 03.65.Pm}
\maketitle

The study of low-momentum scattering in the Schr\"{o}inger equation in
one-dimensional even potentials shows that, as momentum goes to zero, the
reflection coefficient goes to unity unless the potential $V(x)$ supports a
zero-energy resonance\cite{Newton82}. In this case the transmission
coefficient goes to unity, becoming a transmission resonance\cite{Bohm51}.
Recently, this result has been generalized to the Dirac equation\cite%
{Dombey00}, showing that transmission resonances at $k=0$ in the
Dirac equation take place for a potential barrier $V=V(x)$ when the
corresponding potential well $V=-V(x)$ supports a supercritical
state. The conclusion is demonstrated in both special examples as
square potential and Gaussian potential, where the phenomenon of
transmission resonance is exhibited clearly in Dirac spinors in the
appropriate shapes and strengths of the potentials. Except for the
both special examples, the transmission resonance is also
investigated in the realistic physical system. In
Ref.\cite{Kennedy02}, a key potential in nuclear physics is
introduced, and the scattering and bound states are obtained by
solving the Dirac equation in the presence of
Woods-Saxon potential, which has been extensively discussed in the literature%
\cite{Guo02,Petri03,Chen04,Guo03,Alhai01}. The transmission
resonance is shown appearing at the spinor wave solutions with a
functional dependence on the shape and strength of the potential.
The presence of transmission resonance in relativistic scalar wave
equations in the potential is also investigated by solving the
one-dimensional Klein-Gordon equation. The phenomenon of resonance
appearing in Dirac equation is reproduced at the one-dimensional
scalar wave solutions with a functional dependence on the shape and
strength of the potential similar to those obtained for the Dirac
equation\cite{Rojas05}. Recently, Villalba and etal. have discussed
the scattering of a relativistic particle by a cusp
potential\cite{Villa03}. They have solved the two-component Dirac
equation in the presence of a spatially one-dimensional symmetric
cusp potential, and derived the conditions for transmission
resonances as well as for super- criticality. Similarly, they have
also solved the Klein-Gordon equation in the presence of a spatially
one-dimensional cusp potential\cite{Villa07}, and obtained the
scattering solutions in terms of Whittaker functions together with
the condition for the existence of transmission resonances.

Due to the transmission resonance appearing in the realistic physical system
for not only Dirac particle but also Klein Gordon particle as illustrated in
the Woods-Saxon potential as well as the cusp potential, it is indispensable
to check the existence of the phenomenon in some other fields. Considering
that the Hulth\'{e}n potential\cite{Hulth42} is an important realistic
model, it has been widely used in a number of areas such as nuclear and
particle physics, atomic physics, condensed matter and chemical physics\cite%
{Varsh90,Jamee86,Barna87,Richa92}. Hence, to discuss the scattering problem
for a relativistic particle moving in the potential is significant, which
may provide more knowledge on the transmission resonance. Recently, there
have been a great deal of works to be put to the Hulth\'{e}n potential in
order to obtain the bound and scattering solutions in the case of relativity
and non relativity\cite{Chen07}. However, the transmission resonance is not
still checked for particle moving in the potential in the relativistic case.
In this paper, we will derive the scattering solution of the Dirac equation
in the presence of the general Hulth\'{e}n potential, and show the
phenomenon of transmission resonance as well as its relation to the
parameters of the potential.

According to the definition in Ref.\cite{Hulth42,Qiang07}, the general Hulth%
\'{e}n potential is chosen as
\begin{equation}
V(x)=\Theta (-x)\frac{V_{0}}{e^{-ax}-q}+\Theta (x)\frac{V_{0}}{e^{ax}-q},
\end{equation}%
where all the parameters $V_{0}$, $a$, and $q$ are real and positive
together with $q<1$ being required to remove off the divergence of Hulth\'{e}%
n potential, i.e., $0<q<1$ here. If $q=0$, the Hulth\'{e}n potential
will degenerate to the cusp potential represented in
Ref.\cite{Villa03}, where the transmission resonances have been
discussed in details. $\Theta (x)$ is the Heaviside step function.

In order to investigate how the transmission resonance happens in the
potential for a Dirac particle, following a similar procedure to that used
by \cite{Kennedy02}, the Dirac equation takes the form ($\hbar =c=1$)
\begin{equation}
\left[ \gamma ^{\mu }\left( \frac{\partial }{\partial x^{\mu }}-iqA_{\mu
}\right) +m1\right] \Psi =0,
\end{equation}%
where the four-vector potential $A_{\mu }$ can be written in a covariant way
as $qA_{\mu }=iV(x)\delta _{\mu }^{4}$ \ with $\gamma ^{4}=\beta $ and $%
x^{4}=it$ there. When limited to the 1+1 dimensions, the Dirac equation (1)
in the presence of the spatially dependent electric field becomes
\begin{equation}
\left[ \gamma ^{1}\frac{\partial }{\partial x}-i\beta \frac{\partial }{%
\partial t}+\beta V(x)+m1\right] \Psi =0,
\end{equation}%
In Eq.(3), $V(x)$ is independent on time, hence the time dependence of the
spinor $\Psi $ can be separated with $\Psi =e^{-iEt}\psi $ as follows
\begin{equation}
\left[ \gamma ^{1}\frac{d}{dx}-\left( E-V(x)\right) \beta +m1\right] \psi =0.
\end{equation}%
Taking into account that we are working in the 1+1 dimensions, it is
possible to choose the following representation of the Dirac matrices, i.e.,
taking the gamma matrices $\gamma ^{1}$ and $\beta $ to be the Pauli
matrices $\sigma _{x}$ and $\sigma _{z}$ respectively, then the Dirac
equation truns into
\begin{equation}
\left[ \sigma _{x}\frac{d}{dx}-\left( E-V(x)\right) \sigma _{z}+m1\right]
\psi (x)=0.
\end{equation}%
The four-spinor, $\psi (x)$, is decomposed into two spinors, $u_{1}$ and $%
u_{2}$, so that
\begin{equation}
\psi (x)=\left(
\begin{array}{c}
u_{1}\left( x\right)  \\
u_{2}\left( x\right)
\end{array}%
\right) ,
\end{equation}%
Thus the problem is to solve the coupled differential equations:
\begin{eqnarray}
u_{1}^{\prime }\left( x\right)  &=&-\left( m+E-V(x)\right) u_{2}\left(
x\right) , \\
u_{2}^{\prime }\left( x\right)  &=&-\left( m-E+V(x)\right) u_{1}\left(
x\right) .
\end{eqnarray}%
By introducing the following combinations
\begin{equation}
\phi \left( x\right) =u_{1}\left( x\right) +iu_{2}\left( x\right) ,\ \ \ \
\chi \left( x\right) =u_{1}\left( x\right) -iu_{2}\left( x\right) .
\end{equation}%
Substituting these into (7) and (8) and re-arranging gives:
\begin{eqnarray}
\phi ^{\prime }\left( x\right)  &=&-im\chi \left( x\right) +i\left(
E-V(x)\right) \phi \left( x\right) , \\
\chi ^{\prime }\left( x\right)  &=&im\phi \left( x\right) -i\left(
E-V(x)\right) \chi \left( x\right) .
\end{eqnarray}%
The two components, $\phi \left( x\right) $ and $\chi \left( x\right) $,
satisfy:
\begin{eqnarray}
\phi ^{\prime \prime }+\left[ \left( E-V\right) ^{2}-m^{2}+iV^{\prime }%
\right] \phi  &=&0, \\
\chi ^{\prime \prime }+\left[ \left( E-V\right) ^{2}-m^{2}-iV^{\prime }%
\right] \chi  &=&0.
\end{eqnarray}

In order to obtain the scattering solutions for $x<0$ with $E^{2}>1$,
substituting the potential presented in (1) into (12) gives
\begin{equation}
\frac{d^{2}\phi (x)}{dx^{2}}+\left[ \left( E-\frac{V_{0}}{e^{-ax}-q}\right)
^{2}-m^{2}+i\frac{V_{0}ae^{-ax}}{\left( e^{-ax}-q\right) ^{2}}\right] \phi
(x)=0.
\end{equation}%
On making the substitution $y=qe^{ax}$, Eq.(14) becomes%
\begin{equation}
a^{2}y^{2}\frac{d^{2}\phi }{dy^{2}}+a^{2}y\frac{d\phi }{dy}+\left[ \left( E-%
\frac{V_{0}}{q}\frac{y}{1-y}\right) ^{2}-m^{2}+\frac{iV_{0}a}{q}\frac{y}{%
\left( 1-y\right) ^{2}}\right] \phi (x)=0.
\end{equation}%
In order to derive the solution of Eq.(15), we put $\phi =y^{\mu }\left(
1-y\right) ^{\lambda }f$, then Eq.(15) reduces to the hypergeometric
equation
\begin{equation}
y\left( 1-y\right) f^{\prime \prime }+\left[ 1+2\mu -\left( 2\mu +2\lambda
+1\right) y\right] f^{\prime }-\left( 2\mu \lambda +\frac{2V_{0}E}{a^{2}q}%
\right) f=0,
\end{equation}%
where the primes denote derivatives with respect to $y$ and the following
abbreviations have been used
\begin{eqnarray}
\mu &=&\frac{ik}{a},\text{ }\nu =\frac{ip}{a},\text{\ }\lambda =\frac{iV_{0}%
}{aq},  \nonumber \\
p^{2} &=&\left( E+V_{0}/q\right) ^{2}-m^{2},\text{ }k^{2}=E^{2}-m^{2}.
\end{eqnarray}%
Note that as we are considering scattering states, $|E|>m$ which ensures
that $k$ is real, and $V_{0}$ is real and positive. $p$ is real for $q>0$.
The general solution of Eq.(16) can be expressed in terms of hypergeometric
function as
\begin{equation}
f(y)=A\text{ }F\left( \mu -\nu +\lambda ,\mu +\nu +\lambda ,1+2\mu ;y\right)
+B\text{ }y^{-2\mu }F\left( -\mu -\nu +\lambda ,-\mu +\nu +\lambda ,1-2\mu
;y\right) .
\end{equation}%
So
\begin{equation}
\phi _{L}(y)=A\text{ }y^{\mu }\left( 1-y\right) ^{\lambda }F\left( \mu -\nu
+\lambda ,\mu +\nu +\lambda ,1+2\mu ;y\right) +B\text{ }y^{-\mu }\left(
1-y\right) ^{\lambda }F\left( -\mu -\nu +\lambda ,-\mu +\nu +\lambda ,1-2\mu
;y\right) .
\end{equation}%
As $x\longrightarrow -\infty $, there is $y\longrightarrow 0$. So, the
asymptotic behavior of $\phi _{L}(y)$ can be written as
\begin{equation}
\lim_{x\longrightarrow -\infty }\phi
_{L}(x)=Aq^{ik/a}e^{ikx}+Bq^{-ik/a}e^{-ikx}.
\end{equation}%
From equation (10) the other component, $\chi (x)$ is
\begin{equation}
\chi \left( x\right) =\frac{1}{im}\left[ i\left( E-V\left( x\right) \right)
\phi (x)-\phi ^{\prime }(x)\right]
\end{equation}%
Substituting equation (20) into the above gives us
\begin{equation}
\lim_{x\longrightarrow -\infty }\chi _{L}\left( x\right) =A\left( \frac{E-k}{%
m}\right) q^{ik/a}e^{ikx}+B\left( \frac{E+k}{m}\right) q^{-ik/a}e^{-ikx}
\end{equation}%
The choice of combinations of the wave function components (9) can be
re-written :
\begin{equation}
u_{1}=\frac{1}{2}\left( \phi (x)+\chi \left( x\right) \right) ,\text{ \ }%
u_{2}=\frac{1}{2i}\left( \phi (x)-\chi \left( x\right) \right) .
\end{equation}%
Upon substitution of equations (20) and (22) into the above it can be seen
that the wave function, $\psi (x)$, comprises of an incident and reflected
wave far to the left of the barrier which is the desired form to establish
reflection and transmission amplitudes.

Next, we consider the solution of Eq.(3) for $x>0$. With the potential
represented in Eq.(1), the differential equation to solve becomes
\begin{equation}
\frac{d^{2}\phi (x)}{dx^{2}}+\left[ \left( E-\frac{V_{0}}{e^{ax}-q}\right)
^{2}-m^{2}-i\frac{V_{0}ae^{ax}}{\left( e^{ax}-q\right) ^{2}}\right] \phi
(x)=0.
\end{equation}

The analysis of the solution can be simplified making the substitution $%
z=qe^{-ax}$. Eq.(24) can be written as%
\begin{equation}
a^{2}z^{2}\frac{d^{2}\phi }{dz^{2}}+a^{2}z\frac{d\phi }{dz}+\left[ \left( E-%
\frac{V_{0}}{q}\frac{z}{1-z}\right) ^{2}-m^{2}-\frac{iV_{0}a}{q}\frac{z}{%
\left( 1-z\right) ^{2}}\right] \phi (x)=0.
\end{equation}%
Put $\phi =z^{\mu }\left( 1-z\right) ^{-\lambda }g$, Eq.(25) reduces to the
hypergeometric equation
\begin{equation}
z\left( 1-z\right) g^{\prime \prime }+\left[ 1+2\mu -\left( 2\mu -2\lambda
+1\right) z\right] g^{\prime }-\left( -2\mu \lambda +\frac{2V_{0}E}{a^{2}q}%
\right) g=0,
\end{equation}%
where the primes denote derivatives with respect to $z$. The general
solution of Eq.(26) is
\begin{equation}
g(z)=C\left( \mu -\nu -\lambda ,\mu +\nu -\lambda ,1+2\mu ;z\right)
+Dz^{-2\mu }F\left( -\mu -\nu -\lambda ,-\mu +\nu -\lambda ,1-2\mu ;z\right)
.
\end{equation}%
So,
\begin{equation}
\phi _{R}(z)=Cz^{\mu }\left( 1-z\right) ^{-\lambda }F\left( \mu -\nu
-\lambda ,\mu +\nu -\lambda ,1+2\mu ;z\right) +Dz^{-\mu }\left( 1-z\right)
^{-\lambda }F\left( -\mu -\nu -\lambda ,-\mu +\nu -\lambda ,1-2\mu ;z\right)
.
\end{equation}%
Keeping only the solution for the transmitted wave, $C=0$ in Eq.(28). As $%
x\longrightarrow +\infty (z\longrightarrow 0)$, there is
\begin{equation}
\phi _{R}(x)\longrightarrow Dq^{-ik/a}e^{ikx}.
\end{equation}
while the other component
\begin{equation}
\chi _{R}(x)\longrightarrow D\left( \frac{E-k}{m}\right) q^{-ik/a}e^{ikx}.
\end{equation}
The electrical current density for the one-dimensional Dirac equation is
defined as
\begin{equation}
j=\bar{\psi}\left( x\right) \gamma _{x}\psi \left( x\right) =-\psi \left(
x\right) ^{\dagger }\sigma _{2}\psi \left( x\right) =i\left( u_{1}^{\ast
}u_{2}-u_{2}^{\ast }u_{1}\right) =\frac{1}{2}\left( \left\vert \phi \left(
x\right) \right\vert ^{2}-\left\vert \chi \left( x\right) \right\vert
^{2}\right)
\end{equation}%
The current as $x\longrightarrow -\infty $ can be decomposed as $j_{L}=j_{%
\text{in}}-j_{\text{refl}}$ where $j_{\text{in}}$ is the incident current
and $j_{\text{refl}}$ is the reflected one. Analogously we have that, on the
right side, as $x\longrightarrow \infty $ the current is $j_{R}=j_{\text{%
trans}}$, where $j_{\text{trans}}$ is the transmitted current. Using the
reflected $j_{\text{refl}}$ and transmitted $j_{\text{trans}}$ currents, we
have that the reflection coefficient $R$, and the transmission coefficient $%
T $ can be expressed in terms of the coefficients $A$, $B$, and $D$ as
\begin{equation}
R=\frac{j_{\text{refl}}}{j_{\text{in}}}=\frac{\left\vert B\right\vert ^{2}}{%
\left\vert A\right\vert ^{2}}\left( \frac{E+k}{E-k}\right)
\end{equation}
\begin{equation}
T=\frac{j_{\text{trans}}}{j_{\text{in}}}=\frac{\left\vert D\right\vert ^{2}}{%
\left\vert A\right\vert ^{2}}
\end{equation}
Obviously, $R$ and $T$ are not independent; they are related via the
unitarity condition
\begin{equation}
R+T=1.
\end{equation}%
In order to obtain $R$ and $T$ we proceed to equate at $x=0$ the right $\phi
_{_{R}}$ and left $\phi _{_{L}}$ wave functions and their first derivatives.
From the matching condition we can derive the relations among the coefficients $%
A $, $B$, and $D$. Then the reflection coefficient $R$ and
transmission coefficient $T$ are obtained.

The calculated transmission coefficient $T$ varying with the energy
$E$ is displayed in Figs.1-4 at the different values of the
parameters in the Hulth\'{e}n potential. From Figs.1-4, one can see
that the transmission resonance appears in all the Hulth\'{e}n
potential considered here. But the intensity and width of resonance
as well as the condition for the existence of resonance are
different, and they depend on the shape of the potential. Compared
Fig.1 with Fig.2, it can be seen that the width of resonance
decreases as the decreasing of diffuseness $a$, which is similar to
that of Woods-Saxon potential as shown in Figs.3 and 5 in
Ref.\cite{Rojas05}. The same dependence can also be observed from
Figs.3 and 4. Compared Fig.1 with Fig.3, one can find that the
condition for the existence of transmission resonance does also
relate to the parameter $q$. As $q$ decreases, the height of
potential barrier increases, the widths of the transmission
resonance increases. The conclusion can also be obtained by
comparing Fig.2 with Fig.4. In order to obtain more knowledge on the
dependence of transmission resonance on the shapes of the potential,
the transmission coefficient $T$ varying with the strength of
potential $V_{0}$ is plotted in Figs.5 and 6. Beside of the
phenomenon of transmission resonance, similar to the Fig.1 and 2,
the width of resonance decreasing as the decreasing of diffuseness
$a$ is disclosed. All these show the transmission resonances in
Hulth\'{e}n potential for Dirac particle possess the same rich
structure with that we observe in Woods-Saxon potential.

\begin{acknowledgments}
This work was partly supported by the National Natural Science Foundation of
China under Grant No. 10475001 and 10675001, the Program for New Century
Excellent Talents in University of China under Grant No. NCET-05-0558, the
Program for Excellent Talents in Anhui Province University, and the
Education Committee Foundation of Anhui Province under Grant No. 2006KJ259B
\end{acknowledgments}

\begin{figure}[!h]
\centering \includegraphics[width=8.cm]{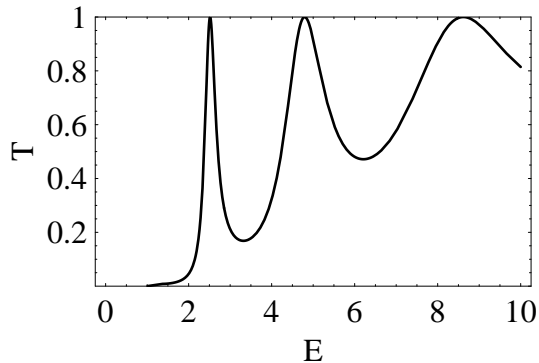} \vspace{-10pt}
\caption{The transmission coefficient for the relativistic Hulth\'en
potential barrier. The plot illustrate $T$ for varying energy $E$ with $%
V_0=4,a=1$, and $q=0.9$.}
\end{figure}
\begin{figure}[!h]
\centering \includegraphics[width=8.0cm]{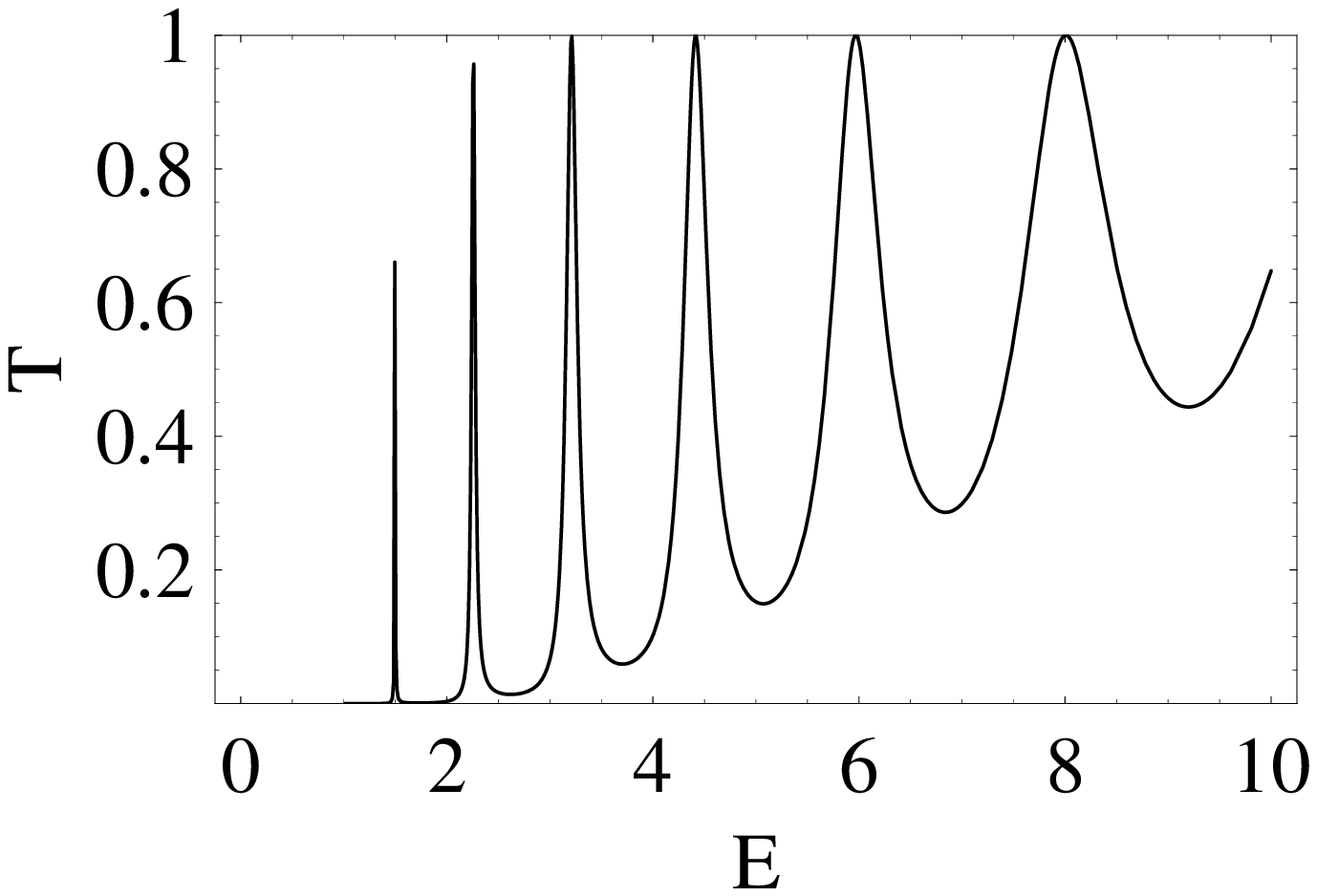} \vspace{-10pt}
\caption{Similar to Fig.1, but with $V_0=4,a=0.5$, and $q=0.9$.}
\end{figure}

\begin{figure}[!h]
\centering \includegraphics[width=8.cm]{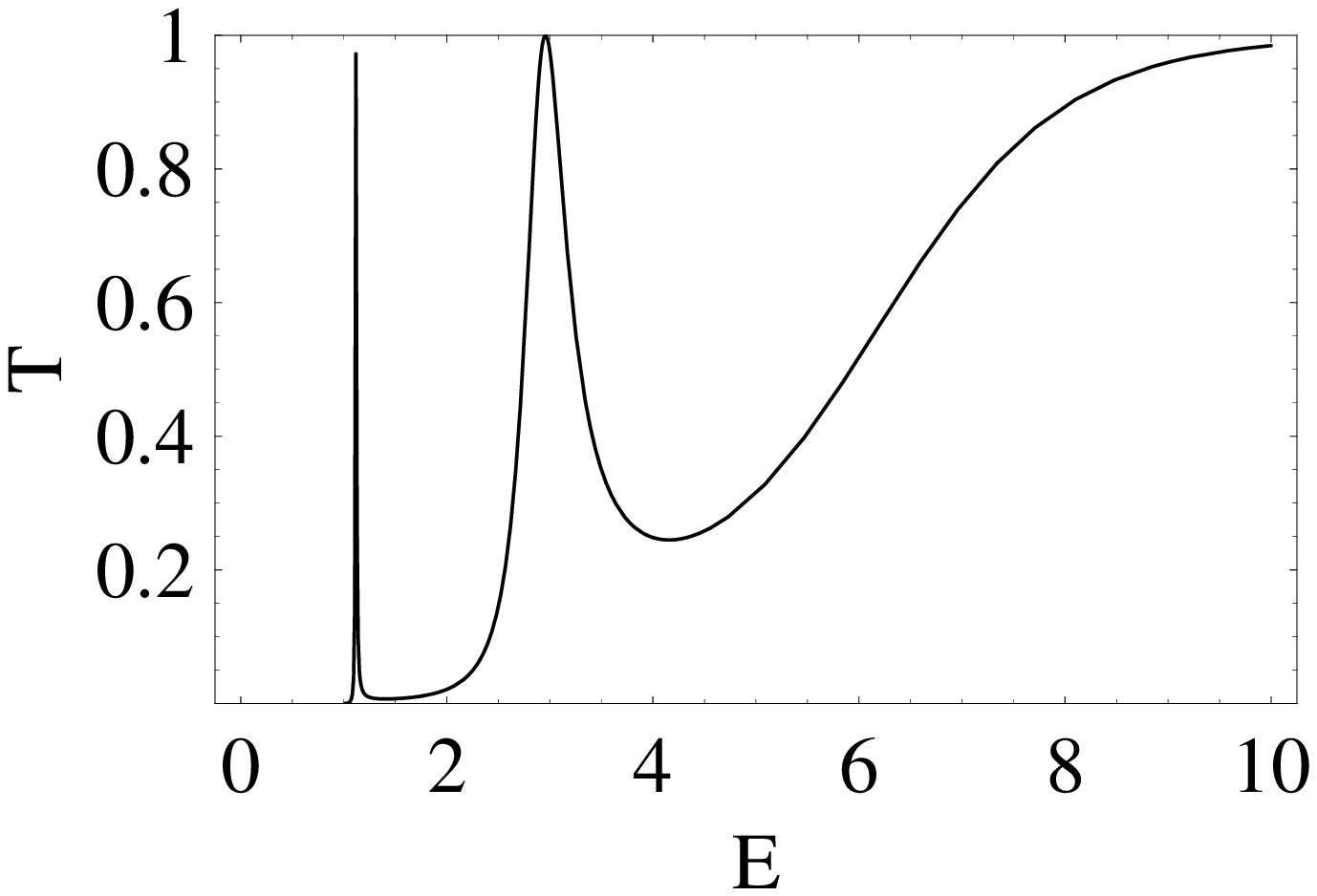} \vspace{-10pt}
\caption{Similar to Fig.1, but with $V_0=4,a=1$, and $q=0.5$.}
\end{figure}
\begin{figure}[!h]
\centering \includegraphics[width=8.0cm]{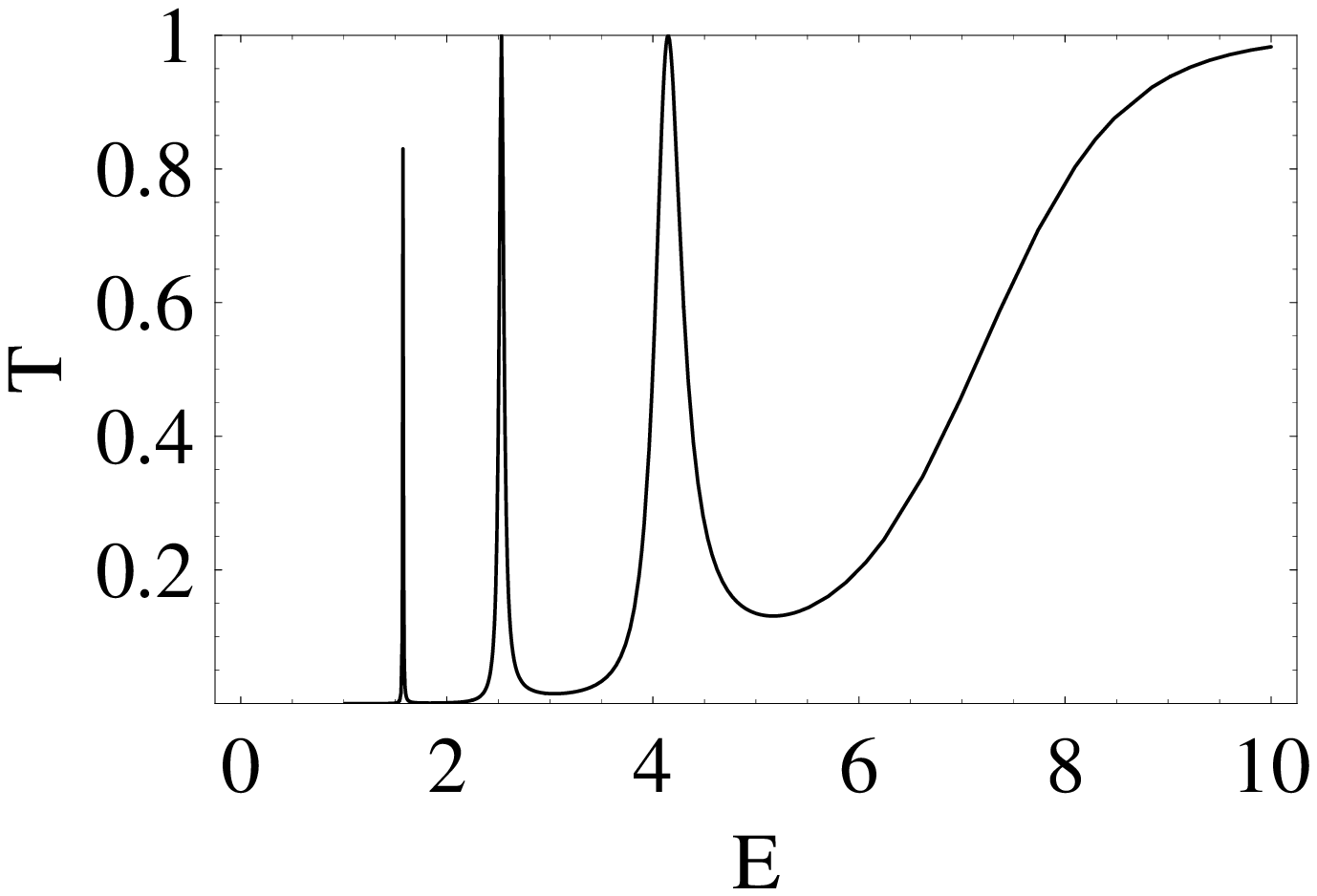} \vspace{-10pt}
\caption{Similar to Fig.1, but with $V_0=4,a=0.5$, and $q=0.5$.}
\end{figure}

\begin{figure}[!h]
\centering \includegraphics[width=8.cm]{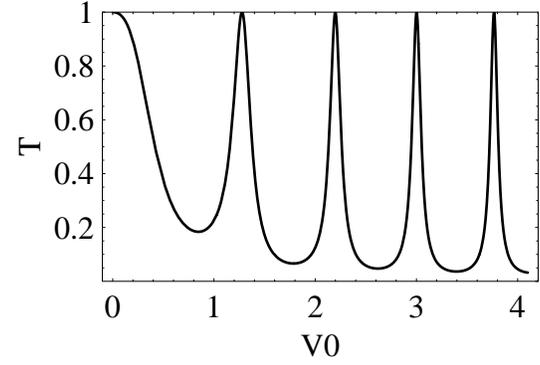} \vspace{-10pt}
\caption{The transmission coefficient for the relativistic Hulth\'en
potential barrier. The plot illustrate $T$ for varying barrier
height $V_0$ with $E=2,a=1$, and $q=0.9$.}
\end{figure}
\begin{figure}[!h]
\centering \includegraphics[width=8.0cm]{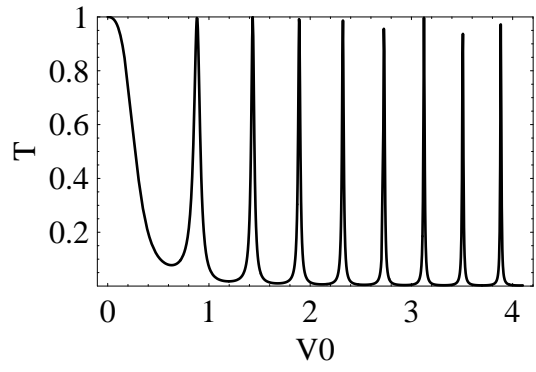} \vspace{-10pt}
\caption{Similar to Fig.5, but with $E=2,a=0.5$, and $q=0.9$.}
\end{figure}

\end{document}